\documentclass[conference]{IEEEtran}
\IEEEoverridecommandlockouts
\usepackage{cite}
\usepackage{amsmath,amssymb,amsfonts}
\usepackage{algorithmic}
\usepackage{graphicx}
\usepackage{textcomp}
\usepackage{xcolor}
\usepackage{url}
\usepackage{multirow} 
\usepackage{braket}
\usepackage{hyperref}
\usepackage{tikz}
\def\BibTeX{{\rm B\kern-.05em{\sc i\kern-.025em b}\kern-.08em
    T\kern-.1667em\lower.7ex\hbox{E}\kern-.125emX}}
\newcommand\submittedtext{
  \footnotesize This work has been submitted to the IEEE for possible publication. Copyright may be transferred without notice, after which this version may no longer be accessible.}

\newcommand\submittednotice{
\begin{tikzpicture}[remember picture,overlay]
\node[anchor=south,yshift=10pt] at (current page.south) {\fbox{\parbox{\dimexpr0.65\textwidth-\fboxsep-\fboxrule\relax}{\submittedtext}}};
\end{tikzpicture}
}
\renewcommand\fbox{\fcolorbox{black}{white}}
\begin{document}

\title{Quantum Architecture Search: A Survey
\thanks{This work was partially funded by the BMWK project EniQmA (01MQ22007A) and and
Einstein Research Unit on Quantum Digital Transformation (ERU-2020-607). \textit{(Darya Martyniuk and Johannes Jung are co-first authors.)}}
}

\author{\IEEEauthorblockN{Darya Martyniuk}
\IEEEauthorblockA{
\textit{Fraunhofer FOKUS} \\
\\
Berlin, Germany \\
darya.martyniuk@fokus.fraunhofer.de}
\and
\IEEEauthorblockN{Johannes Jung}
\IEEEauthorblockA{\textit{Freie Universitat Berlin} \\
\textit{Fraunhofer FOKUS}\\
Berlin, Germany \\
johannes.jung@fokus.fraunhofer.de}
\and
\IEEEauthorblockN{Adrian Paschke}
\IEEEauthorblockA{\textit{Freie Universitat Berlin} \\
\textit{Fraunhofer FOKUS}\\
Berlin, Germany \\
adrian.paschke@fokus.fraunhofer.de}
}

\maketitle
\submittednotice

\begin{abstract}
Quantum computing has made significant progress in recent years, attracting immense interest not only in research laboratories but also in various industries. 
However, the application of quantum computing to solve real-world problems is still hampered by a number of challenges, including hardware limitations and a relatively under-explored landscape of quantum algorithms, especially when compared to the extensive development of classical computing. 
The design of quantum circuits, in particular parameterized quantum circuits (PQCs), which contain learnable parameters optimized by classical methods, is a non-trivial and time-consuming task requiring expert knowledge. As a result, research on the automated generation of PQCs, known as quantum architecture search (QAS), has gained considerable interest. QAS focuses on the use of machine learning and optimization-driven techniques to generate PQCs tailored to specific problems and characteristics of quantum hardware. 
In this paper, we provide an overview of QAS methods by examining relevant research studies in the field.
We discuss main challenges in designing and performing an automated search for an optimal PQC, and survey ways to address them to ease future research.
\end{abstract}


\section{Introduction}
Advancing quantum computing involves developing and improving corresponding hardware and software systems. 
In recent years, research endeavors have led to rapid and impressive progress both in the physical realization of quantum computing concepts and in the development of software tools. 
However, there are still several serious challenges for the research community to overcome before quantum systems can be applied to real-world use cases. 
Machine learning~(ML) and optimization algorithms can be used to open up the potential of hardware devices and expand the possibilities for programming these devices to effectively tackle complex problems. When addressing a particular task with quantum computing, the automation of algorithm design and its execution, including compilation and selection of a suitable device, are promising research directions. Quantum architecture search~(QAS)~\cite{zhang2022differentiable} represents a variety of techniques tailored to automate the process of finding an optimal parametrized quantum circuit~(PQC). 

A PQC is a crucial component of variational quantum algorithms~(VQAs), which have gained great interest in the community and have been successfully applied to various domains, such as chemical simulation~\cite{peruzzo2014variational}, combinatorial optimization~\cite{farhi2014quantum}, and ML~\cite{farhi2018classification,schuld2020circuit}.
It depends on learnable parameters~$\vec{\theta}$, representing angles of quantum rotation gates. 
During the iterative execution of a VQA, a 
classical optimization algorithm adjusts~$\vec{\theta}$ in order to minimize the underlying cost function~\cite{VQACerezo}. 
In PQCs, a subroutine that consist of a sequential application of gates to specific qubits and thus dictates how quantum computation is performed, is called an\textit{ ansatz}.  
Despite the intuitive principle of PQCs, the manual design of a beneficial circuit ansatz is non-trivial, since the criteria for an optimal ansatz in a particular scenario is still an active research area. 
Moreover, PQCs may encounter trainability issues, such as barren plateaus~\cite{cerezo2021cost}. To assists developers in creating powerful PQCs, several manually designed
ansatz patterns have been presented in the literature,~e.g.,~unitary coupled cluster (UCC) and hardware-efficient ansatz~(HEA)~\cite{kandala2017hardware}.
Although the ansatz patterns are useful, they usually need to be tuned in terms of hyperparameters, which is a time-consuming process with no guarantee of success in fitting the particular scenario.
Moreover, quantum noise and hardware constraints further impede the performance of ansatz patterns. To overcome these limitations, QAS methods aim to automatically find the optimal structure and the set of parameters for a PQC, tailored to both the underlying problem and the quantum hardware.

In the recent years, several promising techniques for the automated generation of PQCs have been presented. In this paper, we survey QAS methods and its open challenges. 

 We summarize \textbf{our contributions} as follows:
 \begin{itemize}
   \item We outline the relation of QAS to other fields.
    \item We provide a structured introduction in QAS methods.
    \item We discuss techniques to increase efficiency of QAS.
 \item We introduce ideas for further research direction.  
 \end{itemize}
This work includes six sections.
Section~\ref{Relations} outlines a brief history of QAS and its connection to other fields. 
In Section~\ref{background}, components of QAS are introduced. 
Different search strategies are discussed in Section~\ref{search_strategy}. 
For each strategy, a short overview is given, followed by summaries of selected studies.
In Section~\ref{EI}, we review techniques the improve the efficiency of QAS.
Possible directions for future research are summarized in Section~\ref{ideas}.

\section{QAS Research and Relations to Other Fields} \label{Relations}
\begin{figure}[t]
\centering
    \includegraphics[width=0.9\linewidth]{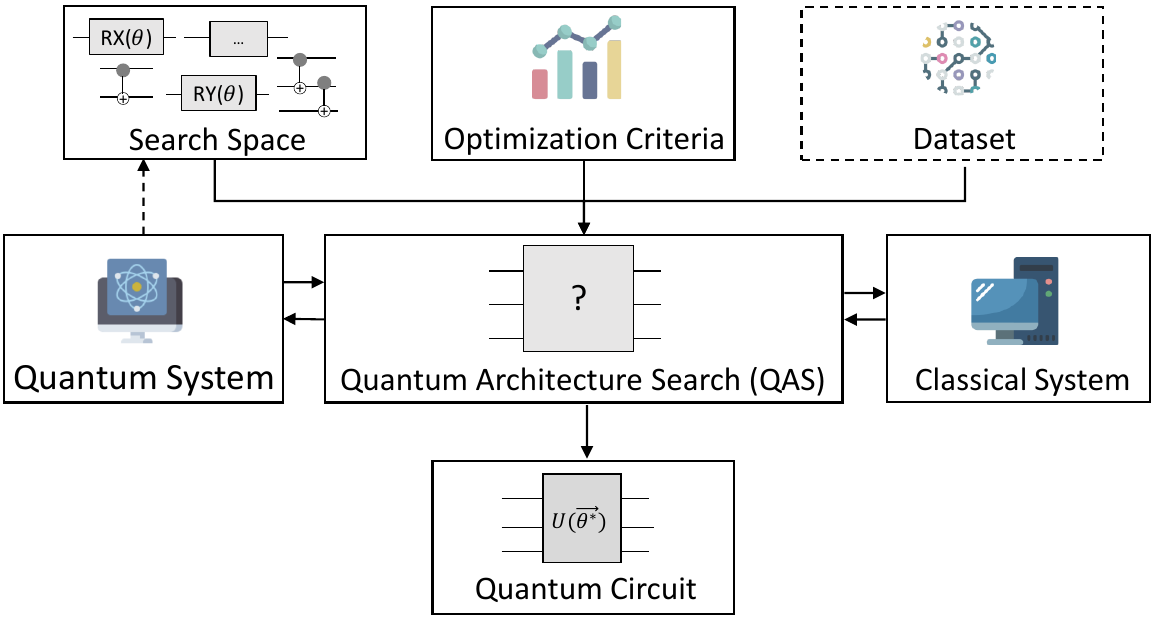}
    \caption{General overview of QAS: Given a search space (which can be constrained by hardware characteristics), performance
criteria, and, in some applications, a dataset, QAS aims to automatically find an optimal PQC with parameters $\vec{\theta^*}$ that maximizes the performance. Quantum and classical systems can be accessed to perform the evaluation of candidate circuits.}  

    \label{fig:QASOverview}
\end{figure}
Initial research into the automatic design of quantum circuits dates back to the late 1990s and early 2000s. 
In early work, researchers used evolutionary algorithms to discover alternative architectures for quantum circuits that are simpler than those constructed manually~\cite{WilliamsColin99,Yabuki2000GeneticAF}, more efficient than classical~\cite{Spector99}, and designed without the need for in-depth knowledge of quantum physics \cite{Rubinstein2001EvolvingQC}.
With the growing accessibility of quantum hardware, simulation environments, and software tools, the interest on the automatic generation of quantum circuits for specific problems has been increased. 
Since then, a wide range of ML techniques such as deep reinforcement learning~\cite{Kuo2021QuantumAS,Zhu:2023sbt, McKiernan2019AutomatedQP, chen2023asynchronous}, differentiable algorithms~\cite{He2021QuantumAS,zhang2022differentiable,pmlr-v202-wu23v,Sun2023DifferentiableQA,He2024-ir}, and Bayesian optimization~\cite{duong2022quantum, pirhooshyaran2021quantum, Nguyen2021QuantumES} have been used to generate or optimize quantum circuits in various application areas, e.g., chemistry~\cite{grimsley2019adaptive,Tang_2021,ryabinkin2020iterative, Bilkis2023-gi, meng2021quantum,fedorov2022unitary, lang2020unitary}, ML~\cite{Lu_2021, altares2021automatic, pmlr-v202-wu23v}, and optimization~\cite{McKiernan2019AutomatedQP, rattew2020domainagnostic}.
The motivation driving the research remains in encoding problems into quantum systems without requiring extensive expertise in quantum computing, designing efficient and noise-resilient circuits in light of limitations of noisy intermediate-scale quantum~(NISQ) hardware, and overcoming algorithmic-specific issues, e.g., trainability issues. 
It should be noted that over the years, the automated design of quantum circuits has been termed in the scientific literature also as quantum circuit design search~\cite{WilliamsColin99, Yabuki2000GeneticAF, pirhooshyaran2021quantum}, quantum circuit search~\cite{Wang2021QuantumNASNS, Anagolum2024livgarEQ}, ansatz architecture search~\cite{li2020quantum}, adaptive variational quantum algorithms~\cite{Turati2023BenchmarkingAV}, quantum circuit learning~\cite{Mitarai_2018, cincio2021machine}, quantum neural architecture search~(QNAS)~\cite{duong2022quantum}, and hybrid optimization~\cite{yao2022monte}. We used these notations to identify relevant studies for this survey. However, some of the publications cover the generation of quantum circuits in general rather than PQCs. Therefore, although these studies have influenced existing QAS methods, they are outside the scope of this survey.

Many QAS techniques introduced in recent years have been strongly inspired by \textit{neural architecture search (NAS)}, which is the process of automated engineering of neural network architectures for a given task. NAS methods have already been successfully applied to a variety of tasks, such as large-scale image classification~\cite{Zoph2016NeuralAS}, segmentation~\cite{Liu2019AutoDeepLabHN}, and text classification~\cite{Xu2021KNASGN}. 
NAS aligns closely with \textit{hyperparameter optimization (HPO)}, which aims to automate the search for optimal hyperparameters of a ML model, i.e., parameters used for the setup of the model or the optimizer (e.g., learning rate, type of optimizer) \cite{Bergstra2013MakingAS}. NAS and HPO methods can be considered as subfields of \textit{automated machine
learning (AutoML)}.
AutoML intends to automate the entire pipeline of a ML model including data preparation and processing, feature engineering, algorithm and architecture selection (e.g., with NAS), and HPO. 
In contrast to related classical fields, QAS extends the idea of architecture search beyond (Q)ML models because of the variety of VQA applications. 
Moreover, due to the current high impact of the concrete hardware on the success of a circuit execution, a lot of QAS studies take into account hardware characteristics developing hardware-aware QAS methods~\cite{Tang_2021,kandala2017hardware,rattew2020domainagnostic,Wang2021QuantumNASNS}. 
The resilience of QAS solutions to noise is another crucial property in the current NISQ-era. 
It has been shown that the presence of noise causes 
the estimated values of the cost function for given parameters to differ from those in a noiseless environment~\cite{Patel2024CurriculumRL}. 
Thus, a number of QAS approaches~\cite{cincio2021machine,Wang2021QuantumNASNS,Patel2024CurriculumRL} define noise resilience as an important property of candidate circuits.

Several works survey techniques and breakthroughs in automated generation of quantum circuits. Reference~\cite{gepp2009review} provide a survey on evolving quantum algorithms using genetic programming.  However, as VQAs had not yet been introduced at the time the work was created, the automated generation of PQCs is not taken into account here. A brief overview of QAS techniques with emphasis on search strategies is provided in~\cite{Zhu_Survey}. Some of QAS methodologies are outlined in~\cite{qin2023review} in the context of a broader review of ansatz designing techniques. 
However, given the active research effort on this topic and novel ideas offered in the recent years, there is a need of a comprehensive overview of QAS methods and challenges. 

\section{Background} \label{background}
\begin{figure}[t]
\centering
\includegraphics[width=1\linewidth]{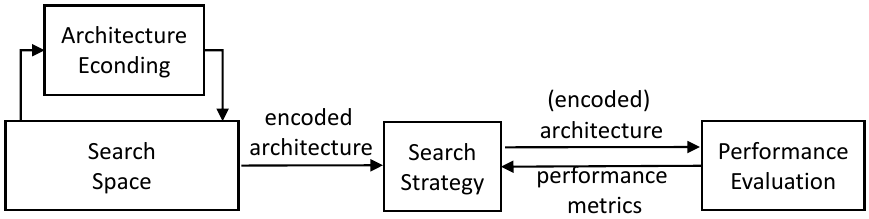}
    \caption{Dimensions of QAS methods. Inspired by \cite{elsken2019neural}.} 
    \label{fig:NASClassification}
\end{figure}

In this work, we define \textit{QAS} as follows. 
Given a task to be solved, QAS aims to automatically design a PQC optimized against specified performance criteria.
Besides to the \textit{task} that specifies the search objective, the process of QAS can incorporate input \textit{data}, e.g., in case of quantum ML applications, or additional \textit{constraints}, e.g., on properties of quantum hardware such as type of supported gates, number of qubits, or qubit connectivity, as illustrated on Fig.~\ref{fig:QASOverview}.
QAS methods can be delineated along four dimensions depicted on Fig.~\ref{fig:NASClassification}.
Note that this categorization is heavily inspired by NAS-related surveys \cite{elsken2019neural,White2023NeuralAS}. 

\textit{\textbf{Search space}} specifies the set of potential architectures that can be discovered.
Due to the large search space, it is typical for QAS studies to constrain it, e.g., by manually constructing the set of gates available for the circuit design. The constraints can be derived from the authors' experience or task's requirements. 
Alternatively, the constrained gate set may be derived from hardware properties, such as the native gate set supported by a particular quantum machine. 

Depending on the search space definition, different levels of granularity are available for the circuit generation.
The search strategy can have access to a pool of individual gates and seek for an optimal arrangement of those to form a PQC.
In such \textit{gate-wise search spaces}, a selected gate can usually be applied to any position as long as it is not occupied by another gate.
However, some QAS approaches, e.g.,~\cite{Sun2023DifferentiableQA}, constrain the working range of gates such that a specific gate can be applied only to dedicated qubits.
Alternatively, a  \textit{layer- and block-wise search spaces} can be used, where circuits are generated by sampling an layer, i.e., a set of gates applied across all qubits simultaneously, or a block of quantum gates, i.e., a set of gates applied 
to one or multiple qubits.
Note that some advanced techniques, such as one-shot methods (see Section \ref{EI}), enable circuit generation by sampling an entire circuit. 

\begin{figure*}[tb]
\begin{center}  \includegraphics[width=1\textwidth]
{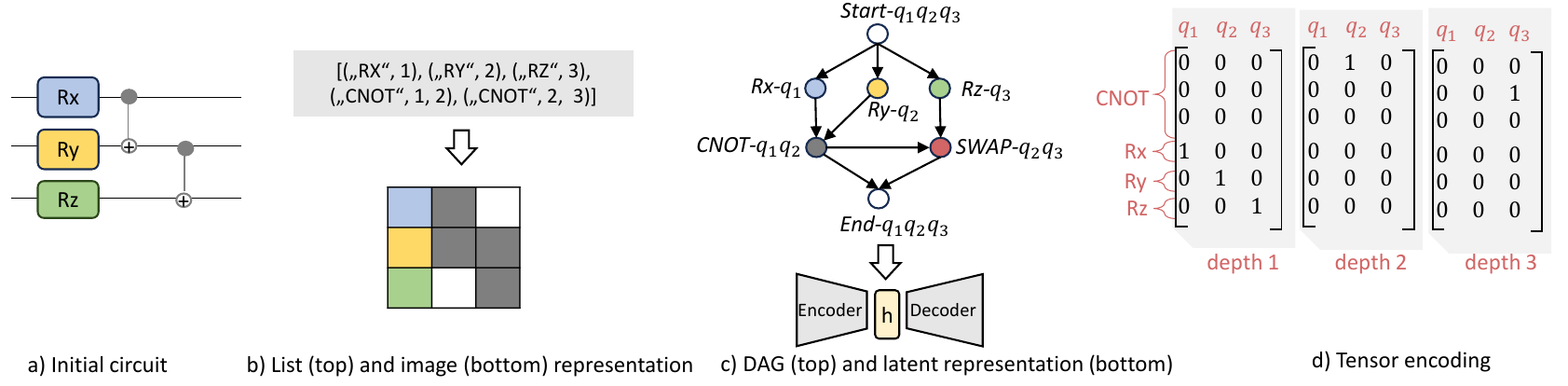}
    \caption {Examples of quantum circuit representations.}
    \label{fig:Encodings}
    \end{center}
\end{figure*}

\textit{\textbf{Encoding techniques}} describe the way of representing quantum circuits within the search space. 
The choice of a circuit encoding significantly impacts the exploration and the manipulation of the search space.
Fig.~\ref{fig:Encodings} illustrates schematically examples of several encoding techniques presented in the literature.
A popular representation for a quantum circuit in QAS is as a \textit{list}~\cite{zhang2021neural,du2022quantum,altares2021automatic} containing elements that encode quantum gates along with their metadata such as positional information. Reference~\cite{zhang2021neural} propose an \textit{image} version of a list encoding, which represents gates and their positions as pixels and can be evaluated by a neural network, e.g., a convolutional neural network~(CNN). Although a list-based encoding offers simplicity in understanding and implementation, it may require assumptions that artificially restrict the search space, e.g., that a quantum machine yields bidirectional
nearest neighbor qubit connections only, i.e.,  $qubit_0 \leftrightarrow qubit_1 \leftrightarrow ... \leftrightarrow qubit_n  $, or that two-qubit gates are symmetric. In addition, decoding a list-based circuit representation can be ambiguous.

Representation of a PQC as a graph is also widely employed in the literature.  
A \textit{directed acyclic graph~(DAG)} provide a compact representation of a quantum circuit, where nodes correspond to quantum gates and directed edges express input and output dependencies~\cite{Nam2017AutomatedOO,duong2022quantum,He2023AGP,He2023GSQASGS,He2024-ir, Sun2024QuantumAS}. 
A \textit{directed multi-graph} representation is used by~\cite{Nguyen2021QuantumES} to encode entangling structures. References~\cite{He2023GSQASGS,Sun2024QuantumAS} propose learning of an \textit{unsupervised representation} of a quantum circuit expressed as DAG via an autoencoder.
In \cite{Lu_2021,Rosenhahn2023MonteCG}, quantum circuits are encoded as paths in (not necessarily acyclic) graphs, whereas in Monte Carlo tree search (MCTS) methods they are encoded as paths in a tree \cite{meng2021quantum,wang2023automated, yao2022monte}.
Graph-based encoding methods allow a compact representation of a quantum circuit. 
However, the may be limited in the expression of asymmetric two-qubit gates such as the CNOT gate.

A quantum circuit can also be expressed as a \textit{tensor}.
A binary encoding scheme introduced in~\cite{Patel2024CurriculumRL}, represents a PQC of maximal ansatz depth $D_{max}$ as a tensor of dimension $[D_{max} \times ((N+3) \times N)]$, where $N$ stands for the size of the problem.  
This encoding scheme enables all-to-all qubit connectivity while it is possible to restrict it to  unidirectional nearest neighbor connections $qubit_0~\rightarrow~qubit_1~\rightarrow~... \rightarrow~qubit_n $.  In~\cite{Patel2024CurriculumRL}, the binary encoding scheme was extended to a 3D tensor-based grid circuit representation, with dimensions representing gate type (height of the tensor), qubit index (width of the tensor), and circuit depth.

\textbf{\textit{Search strategy}} encompasses an approach used to explore the search space. In Section~\ref{search_strategy}, we overview evolutionary algorithms, reinforcement learning, differentiable algorithms, Bayesian optimization, adaptive algorithms, Monte Carlo tree methods, and generative models. However, a combination of different methods can be used, resulting in a hybrid search.

\textbf{\textit{Performance evaluation}} describes the process of assessing the performance of candidate circuits in order to get feedback for the search strategy. A naive approach for evaluating candidate architectures is to first train them from scratch and then calculate the ground truth performance by executing them in a noiseless simulation. 
However, recent studies~\cite{Bilkis2023-gi}, \cite{Wang2021QuantumNASNS}, \cite{Patel2024CurriculumRL} have highlighted the importance of the evaluation in the presence of quantum physical noise
due to adverse affects it can produce, e.g., increased trainability challenges caused by noise-induced barren plateaus~\cite{Wang2020NoiseinducedBP}, and increased complexity~\cite{Fontana2022nontrivial}. 
Yet, evaluating circuits by training and executing them in a noisy environment each time the feedback is needed to drive the search, consumes a significant amount of resources.  
Thus, a number of works use performance estimators to reduce the computational costs of the evaluation. 
We delve into some of these techniques in Section~\ref{EI}

\section{Search Strategies} \label{search_strategy}
In this section, we give a brief overview on search strategies applied in QAS works. Table~\ref{table_strategies} categorizes studies covered in this survey by the search strategy and the underlying task.
\begin{table} [!ht]\label{table_strategies}
\caption{Search strategies and tasks in QAS studies}
\begin{center}
\begin{tabular}{|c|c|c|}
\hline
\textbf{\begin{tabular}[c]{@{}c@{}}Strategy\end{tabular}}                     & \textbf{Task}                                   & \textbf{Reference}                        \\ \hline
\multirow{4}{*}{\begin{tabular}[c]{@{}c@{}}Reinforcement \\ Learning\end{tabular}} 
                                        & \begin{tabular}[c]{@{}c@{}}Combinatorial \\ Optimization (QAOA)\end{tabular} & \begin{tabular}[c]{@{}c@{}} \text{\cite{McKiernan2019AutomatedQP} }\end{tabular}                   \\ \cline{2-3} 
                                        &
                                        \begin{tabular}[c]{@{}c@{}}Ground State\\ Approximation (VQE)\end{tabular}                       & \text{\cite{NEURIPS2021_97244127}, CRLQAS~\cite{Patel2024CurriculumRL}}  \\ \cline{2-3}
                                         &  Diagonalization (VQSD) & RL-VQSD~\text{\cite{Kundu2023EnhancingVQ}}
                                          \\ \hline \hline
\multirow{4}{*} {\begin{tabular}[c]{@{}c@{}}Evolutionary \\ Algorithms \end{tabular}}                         
&    \begin{tabular}[c]{@{}c@{}}Ground State\\ Approximation (VQE)\end{tabular}      &

\begin{tabular}[c]{@{}c@{}}\text{QuantumNAS~\cite{Wang2021QuantumNASNS}},\\  \text{EVQE \cite{rattew2020domainagnostic}},\\ 
MoG-VQE \cite{chivilikhin2020mog},\\
QCEAT \cite{huang2022robust}
\end{tabular} \\ \cline{2-3}
 & Classification                                                                 &        \begin{tabular}[c]{@{}c@{}}\text{QuantumNAS~\cite{Wang2021QuantumNASNS}},\\
 \text{MQNE~\cite{Lu_2021}, \cite{altares2021automatic}} \end{tabular}
 
\\ \hline \hline

\begin{tabular}[c]{@{}c@{}}Generative \\ Models\end{tabular}                                &   \begin{tabular}[c]{@{}c@{}}Ground State\\ Approximation (VQE)\end{tabular}                  &\text{GQE~\cite{Nakaji2024TheGQ}} \\ \hline \hline
\multirow{2}{*}{\begin{tabular}[c]{@{}c@{}}Random \\ Search \\
(with One-Shot\\or Predictor\\
Evaluation)\end{tabular}}
& Classification
& \begin{tabular}[c]{@{}c@{}} \text{\cite{du2022quantum}},\text{\'Eliv{\'a}gar~\cite{Anagolum2024livgarEQ}},\\ \text{\cite{zhang2021neural}},
\text{GSQAS~\cite{He2023GSQASGS}},\\
\text{\cite{Linghu2022QuantumCA}}
\end{tabular}  
\\ 
\cline{2-3}
& \begin{tabular}[c]{@{}c@{}}Ground State\\Approximation \\(VQE)\end{tabular}                                                 & \begin{tabular}[c]{@{}c@{}} \text{\cite{du2022quantum}, TF-QAS~\cite{He_Deng_Zheng_Li_Situ_2024}},\\ \text{\cite{zhang2021neural}},\text{GSQAS~\cite{He2023GSQASGS}},\\
\text{PQAS-AL~\cite{deng2023progressive},\cite{situ2024distributed}}

\end{tabular} 
\\
\hline \hline
\multirow{3}{*}
{\begin{tabular}[c]{@{}c@{}}Differentiable\\ QAS\end{tabular}}
& Noise Reduction                                                  & \begin{tabular}[c]{@{}c@{}}\text{DQAS \cite{zhang2022differentiable}},\\
\text{QuantumDARTS \cite{pmlr-v202-wu23v}} \end{tabular} \\ \cline{2-3}

&  \begin{tabular}[c]{@{}c@{}}Ground State\\ Approximation (VQE)\end{tabular}                                                               &        \text{ QuantumDARTS \cite{pmlr-v202-wu23v}}  \\  \cline{2-3}
& Classification                                                               &        \text{ QuantumDARTS \cite{pmlr-v202-wu23v}}  \\  \cline{2-3}
& \begin{tabular}[c]{@{}c@{}}Combinatorial \\ Optimization (QAOA)   \end{tabular}                                                             &      \begin{tabular}[c]{@{}c@{}}   \text{DQAS \cite{zhang2022differentiable}},\\ \text{QuantumDARTS \cite{pmlr-v202-wu23v}}  \end{tabular}  \\ \cline{2-3}
& \begin{tabular}[c]{@{}c@{}}Quantum\\Reinforcement Learning\end{tabular}                                                               & \text{DQAS-RL\cite{Sun2023DifferentiableQA}}
\\ \cline{2-3}
& \begin{tabular}[c]{@{}c@{}}Compilation\end{tabular}                                                               & \text{MetaQAS\cite{He_2021}, \cite{He2024-ir}}
\\
\hline \hline
\multirow{4}{*}{\begin{tabular}[c]{@{}c@{}}Bayesian \\Optimization                       \end{tabular}}                                        & {\begin{tabular}[c]{@{}c@{}}Quantum Program\\Synthesis                         \end{tabular}}
                  &  \text{\cite{duong2022quantum}} \\\cline{2-3}
& \begin{tabular}[c]{@{}c@{}}Combinatorial \\ Optimization (QAOA)\end{tabular}                          &  \text{\cite{duong2022quantum}} \\ \cline{2-3}
& \begin{tabular}[c]{@{}c@{}}Distribution \\Approximation \end{tabular}                          &  \text{\cite{duong2022quantum}}     \\ \cline{2-3}
& Classification                          &  \text{\cite{pirhooshyaran2021quantum}, QES~\cite{Nguyen2021QuantumES}}  
\\ \hline
\hline
\multirow{3}{*}{\begin{tabular}[c]{@{}c@{}}Adaptive\\Methods \end{tabular}}  
    & {\begin{tabular}[c]{@{}c@{}}Quantum Data\\ Compression \\ (Quantum Autoencoder)                         \end{tabular}}
                  &  \text{VAns~\cite{Bilkis2023-gi}} \\\cline{2-3}
                  
    & \begin{tabular}[c]{@{}c@{}}Ground State\\ Approximation (VQE) \end{tabular} 
        &  \begin{tabular}[c]{@{}c@{}} 
          ADAPT-VQE \cite{grimsley2019adaptive},\\\text{VAns~\cite{Bilkis2023-gi}},\\ \text{Rotoselect \cite{Ostaszewski_2021}},
      \end{tabular}    \\ \cline{2-3}
        
    &Compilation
        &  
         \begin{tabular}[c]{@{}c@{}}  \text{VAns~\cite{Bilkis2023-gi}},\\ \text{QAQC~\cite{khatri2019quantum}}, \\
                 \text{NACL~\cite{cincio2021machine}}
                  \end{tabular}

        \\ \cline{2-3}

    & \begin{tabular}[c]{@{}c@{}}Combinatorial\\ Optimization (QAOA) \end{tabular}  
        & \begin{tabular}[c]{@{}c@{}} AAS~\cite{li2020quantum}  \end{tabular}  \\ \cline{2-3}

    & \begin{tabular}[c]{@{}c@{}}Quantum Program\\ Synthesis \end{tabular}  
        & \begin{tabular}[c]{@{}c@{}} \cite{Cincio_2018}  \end{tabular} \\\cline{2-3} 
      & \begin{tabular}[c]{@{}c@{}}Observable Extraction \end{tabular}                         &  \text{NACL~\cite{cincio2021machine}
}
\\\cline{2-3} 
& \begin{tabular}[c]{@{}c@{}}State Preparation \end{tabular}                         &  \text{NACL~\cite{cincio2021machine}
}
\\ \hline \hline
\multirow{3}{*}{\begin{tabular}[c]{@{}c@{}}Monte-Carlo\\Tree Search                        \end{tabular}}                                        

& \begin{tabular}[c]{@{}c@{}}Error Detection \end{tabular}                         &  \text{~\cite{wang2023automated}
}
\\\cline{2-3}
& \begin{tabular}[c]{@{}c@{}}Linear Equations \end{tabular}                         &  \text{~\cite{wang2023automated}
}
\\\cline{2-3}
& \begin{tabular}[c]{@{}c@{}}Ground State \\ Approximation (VQE) \end{tabular}                         &  \text{~\cite{meng2021quantum}, \cite{wang2023automated}
}
\\\cline{2-3}
& \begin{tabular}[c]{@{}c@{}}Combinatorial\\ Optimization (QAOA) \end{tabular}  
        & \begin{tabular}[c]{@{}c@{}} \cite{wang2023automated},\\
        \text{MCTS-QAOA~\cite{yao2022monte}}  \end{tabular}  \\ \cline{2-3}
& \begin{tabular}[c]{@{}c@{}}Quantum Program\\ Synthesis \end{tabular}  
        & \begin{tabular}[c]{@{}c@{}} MCGS~\cite{Rosenhahn2023MonteCG}  \end{tabular}  \\ \cline{2-3}
& \begin{tabular}[c]{@{}c@{}} Classification \end{tabular}  
        & \begin{tabular}[c]{@{}c@{}} MCGS~\cite{Rosenhahn2023MonteCG}  \end{tabular}  \\ \cline{2-3}
& \begin{tabular}[c]{@{}c@{}} Cellular Automation \end{tabular}  
        & \begin{tabular}[c]{@{}c@{}} MCGS~\cite{Rosenhahn2023MonteCG}  \end{tabular}
\\ \hline
\end{tabular}
\end{center}
\end{table}

\subsection{Reinforcement Learning}
Many studies take advantage of deep reinforcement learning (Deep RL) to find a beneficial quantum circuit~\cite{ye2021quantum, Kuo2021QuantumAS, Zhu:2023sbt,NEURIPS2021_97244127,Patel2024CurriculumRL}. 
These QAS approaches model the problem as a Markov Decision Process (MDP). MDP provides a formalization of sequential decision making and the learning from interaction to achieve a goal. 
In this setup, the learning system 
is called an \textit{agent}. The agent interacts continually with the \textit{environment} by selecting \textit{actions} based on the current \textit{state} of the environment.
The environment responds by informing the agent of the new \textit{state} and by giving the \textit{reward}, which is a numerical value describing how "good" the performed action was. 
In the RL, the goal of the agent is to find an \textit{optimal policy}~$\pi_{*}$ - a policy that maximizes the cumulative reward it receives over time~\cite{Sutton1998}.
Most RL approaches model \textit{actions} as encoded quantum gates/layers/blocks and \textit{states} as encoded variants of quantum circuits, occasionally enriched with problem-specific knowledge. 
The engineering of the reward function depends on the task to be solved by the PQC and has a significant impact on the success of QAS.

Reference~\cite{McKiernan2019AutomatedQP} propose an automated approach to generate quantum circuits for solving combinatorial optimization problems. 
Specifically, the authors focus on three problems: MaxCut, MaxQP and QUBO. 
The RL agent is trained in a customized OpenAI Gym~\cite{OpenAIGym} environment with the Proximal Policy Optimization (PPO) algorithm applied to a shared actor-critic architecture. In the proposed approach, authors define the \textit{action} space as a finite set of discretized rotation gates RZ and RY. The \textit{state} representation incorporates the measurement outcomes from a number of shots complemented by information specific to the given problem instance, such as the graph structure in the case of MaxCut. In each time step, 
the \textit{reward} is calculated as the cost of the expectation value of the Hamiltonian, which is scaled in range 0 to 1. Experiments were executed on the Rigetti's "Quantum Virtual Machine" quantum simulator (ca.~1,700,000 episodes) and the Rigetti Aspen Quantum Processing Unit (QPU) (150,000 episodes). 

In ~\cite{NEURIPS2021_97244127}, Deep RL with feedback-driven curriculum learning is applied to generate an ansatz for the Variational Quantum Eigensolver (VQE). Feedback-driven curriculum learning, referred to as \textit{moving threshold}, allows the difficulty of the learning task to be dynamically and autonomously adjusted based on the current performance of the agent.  In this way, the agent begins to learn a strategy for constructing PQCs for small problems in order to grasp basic concepts and patterns, and as it becomes proficient at handling simpler tasks, the learning examples presented to the agent become more challenging. The proposed curriculum-based RL is extended to the CRLQAS algorithm in \cite{Patel2024CurriculumRL}. 
The agent's ability to construct shorter circuits is facilitated by the \textit{random halting} scheme of determining the total number of actions at the beginning of each episode. Besides further novel concepts such as illegal actions scheme and the speedup of noisy evaluations discussed in Section~\ref{SSR}~and Section~\ref{HANS}, authors analyze the algorithms performance across different noise profiles sources from real quantum devices. 

Reference~\cite{Kundu2023EnhancingVQ} utilizes a Double Deep Q-Network~(DDQN) with $\epsilon$-greedy policy for the quantum state diagonalization task. 
Specifically, the approach aims to find a problem-inspired ansatz for the Variational Quantum State Diagonalization (VQSD)~\cite{Larose2018VariationalQS} algorithm. 
The available actions include $RX$, $RY$, $RZ$ and $CNOT$ gates represented by using the one-hot encoding technique. The dense reward function is engineered to guide the agent to reach the minimum error for a predefined threshold. Experimental results demonstrate that RL-VQSD is able to construct a more efficient ansatz (in terms of smaller number of gates along with the higher accuracy) compared to the standard Linear Hardware Efficient Ansatz (LHEA).

The main drawbacks of Deep RL include its sampling inefficiency, i.e, 
the large number of environmental interactions are required to learn the optimal policy, and its problematic generalization ability, i.e., adapting policies to learn new tasks or interact in a changed environment can be challenging. This means that each time a hardware device is re-calibrated, a QAS algorithm may require expensive retraining. To address this problem,~\cite{ye2021quantum} propose to continually store the information gained during the training to recall it in order to learn to design quantum circuits under different environmental conditions.

\subsection{Evolutionary Algorithms}
Evolutionary algorithms (EA), or genetic algorithms, are search strategies inspired by the 
principles of biological evolution.
EAs work with a \textit{population} of candidate solutions, called \textit{individuals}, that evolve over \textit{generations} (iterations). 
In the case of QAS, each individual represents a specific quantum circuit architecture.
\textit{Genes} are components defining an individual, e.g., layers or gates in a circuit.
Through \textit{mutations} they can be manipulated, exchanged or removed.
The characterization of the genes defines the search space, and influences the behavior of the EA.
Furthermore, every individual can be assigned a \textit{fitness} value via a fitness function, quantifying the quality of the individual.
\textit{Crossover} describes the method of combining genes of different individuals (usually two) to produce a new individual, potentially inhibiting good features of the parents.
The \textit{selection} strategy determines which individuals survive at the end of each generation.
A very general example of an EA iteration might look as follows:
Start with a population of individuals from the previous generation.
Apply crossover and mutation to create new individuals.
Evaluate the fitness of each individual.
Select individuals via the selection strategy to form the next generation.

In \cite{chivilikhin2020mog}, a multi-objective genetic VQE is build, called MoG-VQE, which aims to optimize the ansatz structure while minimizing the two-qubit gate count.
Genes are characterized as blocks composed of one or several CNOT gates and several
rotation operators.
Mutation operations add or delete those blocks to an individual at a random position.
To reduce the time of the parameter optimization, the authors used an EA-based optimization scheme called the Covariance-Matrix Adaptation Evolutionary Strategy \cite{hansen2003reducing}.
In \cite{altares2021automatic}, a multi-objective GA is implemented to search for resource-efficient feature maps for quantum kernels.

Reference \cite{rattew2020domainagnostic} introduces the Evolutionary Quantum Eigensolver (EVQE).
It implements a weight-sharing strategy (see Section \ref{OST}) and initializes the parameters of a rotation gates such
that the gates perform the identity transformation,
yielding to a much faster convergence than random initialization.
Genes are characterized as layers, which only include universal rotations, controlled universal rotations and identity gates.
The algorithm explicitly forbids crossover operations, arguing that combining two circuits does not guarantee that any of the parent properties will be inherited in terms of cost evaluation.
The asexual reproduction strategy provides a similarity metric, defined by the number of mutations to a common  ancestor of two individuals. 
This similarity metric then can be used to encourage diversity within a population by penalizing individuals that are too similar to each other, such that the risk of falling into local optima reduces.

In \cite{Lu_2021}, an algorithm termed Markovian quantum neuroevolution (MQNE) is introduced.
MQNE encodes one-layered circuits into nodes of a directed graph and identifies quantum circuits with paths in that graph.
For the circuit generation, only adjacent controlled $R_x$ gates and universal single-qubit rotation gates are allowed.
A directed edge between two nodes is added if two one-layer circuits do not produce ambiguities when executed successively. Then, an evolutionary search is introduced to find a good path in the resulting graph.

A robust and resource-efficient multi-species EA, Quantum Circuit Evolution of Augmenting Topologies~(QCEAT), is proposed in~\cite{huang2022robust}.
QCEAT maintains multiple subpopulations that initially evolve independently but eventually migrate into each other. 
This approach allows greater diversity and exploration.
To improve the robustness of the individuals, the authors enhance the fitness function with robustness measures.

\subsection{Differentiable QAS}
Differentiable QAS methods relax the discrete search space of quantum architectures into a continuous and differentiable domain.
For the search space $\mathcal{S}=\{s_1, s_2,...,s_n\}$ of quantum architectures, every candidate solution $s \in \mathcal{S}$ is assigned a probability $p_s$($\alpha_s$), depending on a variable $\alpha_s$.
A loss function $L(\alpha)$ has to be defined based on the probability distribution $p(\alpha)=(p_1(\alpha_1),...,p_n(\alpha_n))$ such that it is differentiable in $\alpha$ in some sense.
Thus, the minimum~$\alpha^*~=~\arg\min_{\alpha} L(\alpha)$ can be attained through a gradient descend method.
In the end, the circuit architecture $s$ with the highest probability $p_s(\alpha_s^*)$ is chosen.

In \cite{zhang2022differentiable}, the authors propose a differentiable quantum architecture search (DQAS),  inspired by the NAS method DARTS~\cite{liu2019darts}. Given a predefined operation pool of quantum gates $V = \{V_1,...,V_M\}, \ M \in \mathbb{N}^+$ and a length $N\in \mathbb{N}^+$, the goal is to find an assignment $k = [k_1,...,k_N], \ k_i \in \{1,...,M\}$  so that the resulting circuit $U(k)=V_{k_1} \cdots V_{k_N}$ optimizes the loss function $L(U(k))$. 
The probabilistic model for $k$ is defined by
$P(k,\alpha) = \prod_{i=1}^N p(k_i, \alpha_i)$, with
$p(k_i, \alpha_i)~:=~ \frac{e^{\alpha_{ij}}}{\sum_k e^{\alpha{ik}}}$.
This expression is differentiable in $\alpha$.
Now, if  at each optimization step, a batch of structural parameters is sampled from $P(k,\alpha)$, the final loss function can be defined as
$\mathcal{L}~=~\sum_{k\mathtt{\sim}P(k,\alpha)} L(U(k,\vec{\theta}))$.
Here, $\vec{\theta}$ stands for the gate parameters in the circuit $U(k)$.
This loss function is differentiable in $\vec{\theta}$, but also differentiable in $\alpha$ through Monte Carlo gradients~(see details in \cite{zhang2022differentiable}).
In~\cite{Sun2023DifferentiableQA}, DQAS is successfully applied to design a PQC for quantum RL.

Contrary, in \cite{pmlr-v202-wu23v}, a different differentiable QAS method named QuantumDARTS is proposed.
Here, a different probabilistic model is chosen and the gradients are calculated through the Gumbel softmax function.

\subsection{Adaptive Algorithms}
"Adaptive" is a rather loose description of algorithms.
Thus, we define it broadly: 
Adaptive algorithms consider an initial circuit architecture and modify one part of it at each stage.
This modification can be random or rule-based.
If the best possible modification is performed at each stage, the search process qualifies as a greedy algorithm.
Adaptive approaches often incorporate problem-specific knowledge 
\cite{li2020quantum, grimsley2019adaptive}. 
Furthermore, since adaptive strategy is a rather general concept, it is often integrated in other search methods.

A greedy algorithm termed Adaptive Derivative-Assembled Pseudo-Trotter
ansatz VQE (ADAPT-VQE) is introduced in \cite{grimsley2019adaptive}.
As usual in VQE, the ground state energy of a molecule approximated.
Starting with a Hartree-Fock state as initial ansatz, fermionic operators are successively applied to the ansatz in a greedy manner.
These fermionic operators represent properties of the molecule and are defined in an operator pool.
At each iteration, the operator with the maximal gradient, indicating the maximal change of energy, is applied to the current ansatz~(see \cite{grimsley2019adaptive} for details).
Due to its success, ADAPT-VQE has become a popular method for VQEs.
In fact, there have been a lot of works that either use similar principles or seek to enhance its methodology \cite{ryabinkin2020iterative, Tang_2021}.
For example, \cite{Tang_2021} proposes qubit-ADAPT-VQE and defines a different operator pool than in \cite{grimsley2019adaptive}, such that the circuit depths can be drastically reduced while ensuring the constructing of an exact ansatz is possible.
However, exploring various ADAPT-VQE-based methods or similar VQE approaches falls beyond the scope of this work.

In \cite{li2020quantum}, a QAS method explicitly for QAOA is developed and used together with a novel Gibbs objective function. 
The strategy of the QAS is to sequentially remove two-qubit gates out of the conventional quantum approximate optimization algorithm~(QAOA) ansatz that corresponds to edges in the given graph, defined by the optimization problem.
The removing of gates is done in a greedy manner.

In \cite{Ostaszewski_2021}, an approach called Rotoselect is used for VQE.
The Rotoselect algorithm follows a predetermined layout of gates, where two-qubit gates are fixed and single-qubit gates get optimized, regarding its generators, in this case Pauli matrices, and parameters simultaneously. 
This can be done in only 7 evaluations per single-qubit gate.
The method can be seen as a light-weight QAS since the overall structure stays the same.
The authors suggest to combine the proposed method with other QAS algorithms, e.g., ADAPT-VQE.

In \cite{Cincio_2018}, a method for finding short-depth circuits is introduced and applied to find a PQC for state overlap, i.e., an alternative SWAP test.
Based on input-output samples of the function to learn, at each iteration, the algorithm modifies the structure of the circuit as well as some post-processing parameter, where smaller updates are more probable.

An adaptive approach  VAns (Variable Ansatz)~\cite{Bilkis2023-gi} is tailored to generate potentially trainable and shallow circuit ansatzes for VQA applications.
The approach combine (i)~the stochastic gate insertion to grow candidate circuits and explore the search space and (ii)~a rule-based classical gate pruning to remove gates and thus simplify the circuit.

Several~\cite{khatri2019quantum, cincio2021machine} use simulated annealing (SA) to modify the initial circuit.
In \cite{khatri2019quantum}, the introduced quantum-assisted quantum compiling (QAQC) method involves the SA-based adaptive construction of a PQC ansatz. After the circuit is modified, gate parameters are re-optimized and evaluated. 
If the new circuit produces a lower cost, the change in the structure is accepted. Otherwise, the change is accepted with a probability that decreases exponentially with the size of the cost difference. 
Similar, in~\cite{cincio2021machine}, SA is employed to generate the structure for noise-resilient quantum circuits. 
\subsection{Monte Carlo Tree Search}
The Monte Carlo tree search (MCTS) \cite{coulom2006efficient} method encodes the search space as a search tree.
In case of QAS, nodes of the tree correspond to components of a quantum circuit, e.g., gates, layers or blocks.
Starting from the root node, a path of the tree builds the circuit by traversing the nodes along the path.
By evaluating different nodes within the search tree, MCTS gathers information about the performance of various circuits. This information is then used to guide the selection process in subsequent iterations.
The selection policy needs to balance exploitation, i.e., selecting previously evaluated, promising nodes, and exploration, i.e., exploring unseen branches.
As shown in \cite{meng2021quantum,wang2023automated}, MCTS methods can be combined with an one-shot strategy~(see Section~\ref{OST}).

Reference~\cite{meng2021quantum} propose a MCTS method for VQEs, where nodes of the Search Tree represent predefined layers.
Selection probabilities of different nodes are determined during the process by the upper confidence bounds for trees \cite{kocsis2006bandit}, a commonly used strategy in MCTS.
In \cite{wang2023automated}, the authors introduce MCTS for a very general class of problems.
They also use upper confidence bounds and combine it with  combinatorial multi-armed bandits \cite{chen2013combinatorial}.
Reference~\cite{yao2022monte} proposes MCTS for QAOA, while~\cite{Rosenhahn2023MonteCG} introduces Monte Carlo graph search, where the structure is changed from a tree to a graph to circumvent ambiguities in the resulting circuits.
\subsection{Bayesian Optimization}
 Bayesian optimization (BO) is a well-known strategy for optimizing black-box functions that are expensive to evaluate. 
 It is used in NAS \cite{kandasamy2018neural}, and first endeavors have been made also for QAS \cite{duong2022quantum, Nguyen2021QuantumES,Sun2024QuantumAS}, \cite{pirhooshyaran2021quantum}.
 Given the objective function $f$ to be minimized and the initial set of quantum circuits, BO begins by constructing a \textit{surrogate model} – a probabilistic model, which approximates $f$ based on the data collected from past evaluations. Then, over a sequence of iterations, a new set of quantum circuits is sampled from the search space according to the \textit{acquisition function}~–~a function derived from the surrogate model and used to determine the most promising samples to evaluate. Acquisition function aims to balance the exploration and exploitation trade-off in BO. Next, the new samples are evaluated to determine the ground truth performance, e.g., validation accuracy in a QML task, and the data collected by the evaluation is used to update the surrogate model. A comprehensive overview on BO can be found in \cite{frazier2018tutorial}.

In \cite{duong2022quantum}, the authors employ a Gaussian process (GP) based surrogate model.
To build the model, they introduce a sophisticated distance measure between circuit architectures using optimal transport. 
Expected Improvement (EI), which is optimized by an EA, is used as acquisition function. 
Reference~\cite{Nguyen2021QuantumES} presents the QES algorithm for searching an optimal structure of an entangling layer.
In QES, Tree Parzen Esimator (TPE)~\cite{Bergstra2011AlgorithmsFH} is used as a surrogate model and EI is employed as acquisition function. Also in \cite{pirhooshyaran2021quantum} and \cite{Sun2024QuantumAS}, BO is used next to other approaches as a search strategy.
\subsection{Generative Models}
The Generative Quantum Eigensolver (GQE) introduced in~\cite{Nakaji2024TheGQ} uses classical generative models to generate PQCs for quantum simulation. In particular, the authors focus on the Transformer architecture~\cite{NIPS2017_3f5ee243}, typically used in Natural Language Processing.
However, instead of generating sequences of words, GQE creates sequences of quantum circuits.
To demonstrate the algorithm realization, the authors introduce GPT Quantum Eigensolver (GPT-QE), which is inspired by the decoder-only transformer GPT-2~\cite{Radford2019LanguageMA}, and describe training and, in case of pre-training, knowledge transfer scenarios. 
Note that the algorithm can also be realized by any other model of the family of generative models, such as Generative Adversarial Networks (GANs), Flow models, Restricted Boltzmann Machines, or Autoencoders. In addition, the proposed idea of using generative models to generate and optimize parameterized circuits can be extended to other QAS tasks besides the quantum simulation. 
\section{Efficiency Improvement} \label{EI}
When performing QAS, the iterative sampling and training of individual circuits from scratch results in high computational costs and presents a challenge in scaling up experiments to larger numbers of qubits or circuit depth.
In this section, we overview methods used to improve the efficiency of QAS.
\subsection{One-shot Techniques} \label{OST}
Multiple works~\cite{du2022quantum, Wang2021QuantumNASNS, Linghu2022QuantumCA, wang2023automated, meng2021quantum, Sun2023DifferentiableQA} have adjusted \textit{one-shot} techniques, also known as \textit{weight-sharing},  introduced in NAS, for the quantum domain. 
Related to QAS, one-shot techniques aim to determine parameters of all individual circuit architectures by a single training of a \textit{super-circuit}, which contains the most extensive set of quantum gates within the specified search space. 
A general scheme of one-shot QAS is as follows: At the beginning,  the super-circuit is initialized and fully trained. At each step, individual sub-circuits, i.e.,  circuits encompassing structures included in the super-circuit and inheriting the parameters, are sampled and evaluated. 
Based on the evaluation, best top-\textit{k} performing sub-circuits are selected and re-trained from scratch. The best sub-circuit is finally returned as the solution.
In QuantumNAS~\cite{Wang2021QuantumNASNS}, sub-circuits are additionally trained for a single iteration and updated parameters are transferred back to the super-circuit.  
In~\cite{du2022quantum}, an iterative training process of the super-circuit is employed.
Specifically, the super-circuit is trained by iteratively sampling and training the set of sub-circuits, rather than training from scratch after the initialization.
This approach was also utilized by~\cite{Linghu2022QuantumCA} for the experimental implementation of QAS on a quantum superconducting processor.
In~\cite{wang2023automated} and~\cite{meng2021quantum}, an one-shot strategy is incorporated in a MCTS such that two paths within the tree share the same weights at nodes (i.e., gates or layers) on which they overlap.
Furthermore, in~\cite{rattew2020domainagnostic}, a weight-sharing is included into the EA, such that individuals share weights with their ancestors.
However, it should be mentioned that some issues associated with one-shot NAS might be also relevant for QAS.
First, it is not ensured that super-net training preserves the same ranking as independent sub-net training.
Second, gradient variations obtained by super-net due to learning by various sub-nets can hurt the convergence and performance~\cite{Gong2022NASViTNA}. These possible drawbacks need to be  studied in future research.

\subsection{Performance predictors}
To increase the efficiency of QAS, a number of works use predictors (or proxies) to forecast the performance of architectures and thus avoid costly training of each architecture to obtain its performance.
A distinction can be made between trainable and training-free (or zero-cost) predictors.  

\textbf{Trainable predictors:}
The idea of a predictor guided QAS was introduced in~\cite{zhang2021neural}.
In this work, classical neural predictors, specifically a Recurrent Neural Network~(RNN) regression model and a Convolutional Neural Network~(CNN) classification model, are used to forecast the performance of candidate quantum architectures.
The predictors are trained on a small subset of quantum circuits and their ground truth performance. Notably, while preparing the training dataset, each circuit is evaluated multiple times with different initialization parameters since the loss landscape may have numerous local minima. 
Once the predictors are trained, the CNN-based predictor, which is designed as a binary classifier to distinguish between "bad" and "good" circuits, is used to filter out weak quantum circuit candidates.  The remaining candidates are further evaluated by the RNN-based predictor. 
Finally, chosen circuits are trained and evaluated from scratch and the best performing circuit is selected as the solution.

The predictor proposed in 
\cite{He2023AGP} consist of (i)~an encoder based on Graph Neural Network~(GNN), which maps the discrete circuit structure encoded as DAG into a continuous feature representation, and (ii)~a multi-layer perceptron regressor, which estimates the performance of the quantum circuit according to the representation learned by the encoder. 

In GSQAS,\cite{He2023GSQASGS}, a graph predictor is trained in a self-supervised manner.
The training incorporates (i) the unsupervised learning of circuit encoding from large numbers of unlabeled quantum circuits as a pretext task (similar routine is presented in \cite{Sun2024QuantumAS}) and subsequently (ii) the fine-tuning of the encoder and the performance estimation predictor from a small number of labeled circuits as a downstream task. 
Candidate circuits are represented as DAGs, enabling the use of Variational Graph Autoencoder (VGAE)
with Graph Isomorphism Networks (GINs)
to learn a latent representation of a quantum circuit that is invariant to isomorphic graphs. 
In \cite{He2022SearchSP}, the authors use a performance estimation for search space pruning, which compares rotation angles after just a few optimization steps.

One drawback of trainable predictors is that the circuit space expands exponentially as the size of quantum circuits increases.
Addressing this problem, \cite{deng2023progressive} introduce PQAS-AL algorithm, which incorporates a progressive predictor training with the active learning technique. 
Active learning enables the learner to select the samples it wishes to learn from, e.g., by querying specific samples to be labeled from an unlabeled data pool. In the initialization phase of PQAS-AL, a large number of unlabeled circuits $C_{u_{0}}$ are sampled from the search space as a basis for the construction of an initial labeled dataset $D_{u_{0}}$. The predictor is then trained on $D_{u_{0}}$. Next, the trained predictor selects a new set of unlabeled quantum circuits $C_{u_{1}}$ with high predicted performance. The ground truth for each circuit in $C_{u_{1}}$ is calculated and the labeled dataset gets updated.The process repeats until the algorithm converges or reaches the maximum number of iterations.

\textbf{Training-free predictors:} 
Another approach to address potentially inaccurate predictors due to the large search space is using training-free predictors.  
In {\'E}liv{\'a}gar~\cite{Anagolum2024livgarEQ}, two training-free predictors are leveraged: the \textit{Clifford noise resilience (CNR) predictor} to filter out low-fidelity architectures in the early stage of the search process and the \textit{representational capacity (RepCap)} predictor to forecast the performance of architectures on the target task. 
TF-QAS~\cite{He_Deng_Zheng_Li_Situ_2024, situ2024distributed} also uses a combination of two training-free predictors to avoid time-consuming training during circuits evaluation. The \textit{path-based} predictor, which is defined as the number of distinct paths between input and output nodes of a DAG and thus can approximate the topological complexity of a circuit, is employed to filter out unpromising architectures. 
However, the number of paths has a relatively weaker correlation with circuit performance. 
For this reason, the \textit{expressibility-based} predictor is used to find high-performance architectures.
In \cite{li2020quantum}, an estimation method is proposed, specific for estimating the performance of QAOA ansatzes produced by the algorithm.
\subsection{Meta-Learning}
Meta-learning, or "learning to learn", is a ML technique that leverages prior experience with other related tasks to help a model to acquire new skills easier, requiring fewer examples or less interactions with an environment~\cite{Vanschoren2018MetaLearningAS}. 
Meta-learning methods such as warm start, transfer learning, active learning, and few-shot learning are widely integrated into the training of (Q)MLs models. 
Reference~\cite{He2021QuantumAS} presents MetaQAS algorithm, which employs a meta-learning on differentiable QAS to learn a good initialization heuristics for quantum circuit architectures, i.e., meta-architecture. 
A significant improvement of running time on a differentiable QAS has been shown in~\cite{He2024-ir}, attributed to a meta-trained generator, which efficiently produces a set of candidate architectures, and a meta-predictor, which evaluates the architectures. In~\cite{zhang2021neural,Nakaji2024TheGQ}, and~\cite{duong2022quantum}, the authors employ a transfer learning methodology to leverage knowledge previously acquired by the model.
\subsection{Search Space Reduction} \label{SSR}
One possibility to facilitate the search of ansatzes is to reduce the search space.
For example, the authors of \cite{He2022SearchSP} propose a search space pruning as preprocessing for QAS.
Given a PQC of length $N$ and candidate gate sets $\mathcal{G}_i$ for every gate position $i$ in the circuit, $\mathcal{G}_i$ are reduced iteratively, decreasing the search space.
This is done by sampling a batch of circuits and removing the candidates out of $\mathcal{G}_i$ that on average perform the worst.
In the context of RL,~\cite{Patel2024CurriculumRL}~introduces the scheme of illegal actions (IA), which cancels out actions that  either reverse the effect of previous actions or are redundant. 
Furthermore, instead of searching for a full PQC ansatz, some studies investigate reduced scenarios. In~\cite{duong2022quantum} and \cite{pmlr-v202-wu23v}, authors reduce the search space in a way that the QAS algorithm generates smaller layer patterns and constructs the final circuit by repeated application of those patterns.
In~\cite{Nguyen2021QuantumES}, 
the search generates an optimal structure of entangling layout.

\section{Challenges and Further Research Directions} \label{ideas}

In this section, we discuss some fundamental open challenges and potential future directions for the QAS research.

\subsection{Benchmarks and Comparability}
When introducing a novel QAS technique, it is typical to evaluate its performance against a random search, state-of-the-art QAS approaches, or human-designed ansatzes.
Lately, a number of studies presented important findings while benchmarking QAS algorithms for specific applications, e.g., quantum chemistry \cite{Claudino2020BenchmarkingAV, Mukherjee2022ComparativeSO, Saib2023BenchmarkingAQ, costa2021benchmarking} and Quadratic Unconstrained Binary Optimization (QUBO) problems \cite{Turati2023BenchmarkingAV}. 
However, the experiments were conducted using evaluation protocols and criteria outlined within the studies rather than on predefined benchmarks. 
This is due to the lack of comprehensive QAS benchmarks to measure methods performance in a uniform, consistent and reproducible manner. To address this problem, \cite{QAS_Bench}\footnote{\url{https://github.com/Lucky-Lance/QAS-Bench}}  introduces 
\textit{QAS-Bench} with two tasks: (i) quantum circuit regeneration, where an QAS algorithm needs to find a circuit that is equivalent to the target circuit, and (ii) approximating an arbitrary unitary, where unitary is assumed to be unavailable for the QAS algorithm, e.g., as in the case of ML scenarios or algorithmic oracles. 
To facilitate the training and benchmarking of RL-based methods for quantum compilation, the customized OpenAI Gym environment, \textit{QGYM }\footnote{\url{https://github.com/QuTech-Delft/qgym}}, was recently introduced in~\cite{Linde2023qgymAG}.

Still, the QAS community could benefit from known from NAS research \textit{tabular benchmarks}, i.e., benchmarks that provide precomputed scores for all possible architectures in the search space such as NAS-201~\cite{Dong2020NASBench201ET}, and \textit{surrogate} benchmarks, i.e., benchmarks that come with a surrogate model that is capable of predicting performance of any architecture in the search space such as JASH-Bench-201~\cite{Bansal2022JAHSBench201AF}. Benchmarks of this kind could reduce the computational cost when comparing different methods and improve fairness through a consistent definition of the search space and the performance evaluation.  Furthermore, since hardware-awareness and noise-resilience of QAS methods are of great importance
in the current NISQ-era, the design of QAS benchmarks should take into account the evaluation of algorithms performance under noisy conditions and specific hardware characteristics, such as qubit connectivity. 
In this regard, designers of QAS benchmarks should be aware of research efforts in comparing quantum compilation methods at low- (QASMBench~\cite{li2020qasmbench}), high- (application-oriented) (SupermarQ~\cite{Tomesh2022SupermarQAS},  QED-C~\cite{Lubinski2024QuantumAE}), and cross-levels (MQT Bench~\cite{quetschlich2023mqtbench}, Arline~\cite{kharkov2022arline}), and task specific benchmarks, e.g., QML-Benchmarks introduced in~\cite{Bowles2024BetterTC}.

It is also important to mention the checklist of best practices for NAS\footnote{\url{https://automl.org/nas_checklist.pdf}}~\cite{lindauer2020best}, which shall also be followed by QAS researchers. However, the checklist can be further extended to include QAS relevant points, e.g., for reporting results of experiments in noisy environments.
\subsection{Search Scope}
Most of the works reviewed in this survey use QAS to find an ansatz for a PQC, while using a fixed data encoding to embed classical data into the quantum system if necessary, e.g., an angle encoding. However, due to the limited capacity of NISQ devices, the dimension of the data must be reduced before applying data encoding. This might change the hardness and nature of the learning problem as mentioned in~\cite{Bowles2024BetterTC}. 
Additional effort on automated design of efficient techniques to encode classical data into quantum circuits and specifically PQCs could help researchers to consider experiments on more realistic data. In addition, discovering new search spaces, e.g., including not only quantum gates, but also classical concepts such as neural network layers, could help to explore uncovered architectures.
Furthermore, a combination of QAS methods with other techniques, such as error-correction,
could further empower them. Finally, the development of end-to-end tools incorporating different stages of automated design such as data encoding, selection, generation, and execution of PQCs, would fulfill one of the motivational points behind the QAS research. 
\subsection{Hardware-Awareness and Noise-Resilience} \label{HANS}
In contrast to NAS, the problem of QAS is often formalized as a multi-objective optimization problem that requires maximizing the circuit performance while, e.g., reducing its depth, and increasing its robustness to hardware noise.
Multiple studies~\cite{ye2021quantum, Bilkis2023-gi, Patel2024CurriculumRL, Wang2021QuantumNASNS, rasconi2019innovative, cincio2021machine} have highlighted the importance of hardware-awareness and noise-resilience of generated quantum circuits.  
Yet, incorporating these effects into the search space and the architectures evaluation, largely increases the complexity of the problem. Therefore, existing hardware-aware algorithms are either specifically designed for the compilation task, e.g.,~\cite{khatri2019quantum}, or  
incorporate only a part of tasks towards compilation of problem-specific PQCs, e.g., qubit mapping in~\cite{situ2024distributed, Wang2021QuantumNASNS}. 
Further research is necessary to explore efficient problem formulations or the combination of methods operating effectively in ensemble. 
In addition, new techniques to reduce the runtime of experiments incorporating a noise model need to be discovered. 
For example, in~\cite{Patel2024CurriculumRL}, the overhead of performing function evaluations in the simulation of a noisy environment was reduced by applying the Pauli-Transfer Matrix (PTM) formalism that allows to precompute offline the fusion of noise channels with their respective gates rather than recompile them at each step during the training. 
Combined with GPU computation and just-in-time (JIT) compiled functions in JAX~\cite{bradbury2018jax}, the PTM formalism enabled a six-fold improvement in the training cost. 

\subsection{Quantum-specific Methods for Encoding and Evaluation}
As mentioned in Section~\ref{background}, the representation of quantum circuit structures in the search space significantly impacts the efficiency of QAS algorithms. Optimizing and discovering new scalable techniques to encode quantum circuits could enable more efficient exploration of search spaces. 
Moreover, most of the methods tailored to accelerate the evaluation of candidate architectures, such as one-shot and predictor-based techniques, are heavily inspired by NAS research. Further inspection on their power and limitations in the context of QAS is needed. Adjusting these method to quantum domain, similar as made in~\cite{Anagolum2024livgarEQ} with the usage of CNR and RepCap predictors, could increase the performance of these techniques.

\subsection{Concurrency Methods}
The ongoing advancement of quantum hardware and its increasing accessibility suggests that in the near future algorithms capable of parallel execution on quantum hardware may become increasingly crucial.
This is especially important for QAS methods since they involve a lot of circuit executions. 
While some search strategies, such as EAs and asynchronous Deep RL, allow multiple agents to learn independently and in parallel in different copies of the environment, e.g., as realized in~\cite{QRLforQAS},
the development of QAS methods specifically designed for parallel processing needs to be investigated in future research.

A closely related approach involves distributing the search across multiple devices that can compute different parts of the task simultaneously.
A distributed search across  multiple small-scale QPUs interconnected by quantum links is introduced in~\cite{situ2024distributed}. Distributed quantum computing~(DQC) involves, in addition to local gates operating on the qubits of a single QPU (data qubits), also non-local gates affecting pairs of qubits on different
devices (communication qubits).
Thus, distributed QAS need to incorporate the graph representation of the distributed quantum system as an input and take into account local and non-local types of  gates as well as the optimization of implementation methods for non-local gates. 
Although the distributed QAS may enable generation of circuits for large-scale problems, the search complexity is greater than in a single QPU scenario. This need to be further investigated, combined with open issues in the DQC research. 

\section{Conclusion}
The landscape of quantum computing is rapidly evolving. While the potential of quantum systems for solving complex problems is beyond doubt, the designing of quantum algorithms in the current NISQ-era remains challenging. Collaboration across research areas can accelerate the design of quantum methods and lead to a better understanding of the quantum computing paradigm.
In this survey paper, we outlined techniques and research directions for the automated generation of PQCs using ML and optimization methods. Due to the inherent complexity of PQCs and the limitations of NISQ devices, QAS research faces significant time and resource challenges. We summarized established techniques to address the computational overhead and concluded the survey with discussing possible future research directions.
\bibliographystyle{IEEEtran}
\bibliography{IEEEabrv, main}
\end{document}